\documentclass[12pt,english]{article}
\usepackage[T1]{fontenc}
\usepackage[latin9]{inputenc}
\usepackage{setspace}
\makeatletter
\usepackage{aStyle}

\makeatother

\usepackage{babel}
\begin{document}
\begin{doublespace}
\begin{center}
\textbf{\Large{}The Epistemology of Contemporary Physics: Introduction}\vspace{-1.3cm}
\par\end{center}
\end{doublespace}

\begin{center}
Taha Sochi (Contact: ResearchGate)\vspace{-0.4cm}
\par\end{center}

\begin{center}
London, United Kingdom
\par\end{center}

\textbf{Abstract}: This is the first of a series of papers that we
intend to publish about the epistemology of fundamental physics in
its current state. One of the main objectives of these papers is to
improve our understanding of fundamental physics (and modern physics
in particular) from an epistemological and interpretative perspective
(i.e. versus formal perspective). Another main objective is to investigate
and assess the merit of searching for a unified physical theory (the
so-called ``theory of everything'') considering the fact that contemporary
physics is a collection of theories created and developed by different
individuals and groups of scientists in different eras of history
reflecting different levels of scientific, philosophical and epistemological
development and dealing with largely separate physical phenomena and
hence such unification may mean ``stitching together'' an inhomogeneous
collection of theoretical structures which may be clumsy (if not impossible)
at least from an epistemological viewpoint.\vspace{0.3cm}


\clearpage{}

\tableofcontents{}

\clearpage{}

\section{\label{secIntroduction}Introduction}

Contemporary physics is a collection of theories created and developed
by different individuals and groups of scholars in different eras
of history reflecting different levels and stages of scientific, philosophical
and epistemological development. These theories deal with the physical
world as separate phenomena (such as electromagnetic phenomenon and
quantum phenomenon) where each phenomenon has its own collection of
laws, rules, principles, methodologies, concepts, terminologies, and
so on. This simply means that contemporary physics is an inhomogeneous
collection of theoretical structures that may not be compatible with
each other from an epistemological viewpoint since they rest on different
theoretical infrastructures.

Although contemporary physics works fine at the formal and practical
level in most areas, it faces serious challenges at the epistemological
and interpretative level as seen, for instance, in the failure so
far of having a sensible and reliable interpretation for quantum mechanics
which is engulfed by many perplexities and ``paradoxes'' despite
its undeniable success at the formal and practical level (see for
example \cite{Laloe2002,SochiBook8}). This dilemma (in our view)
is not limited to quantum mechanics but it extends to other physical
branches and disciplines that are largely seen as well-understood
and -interpreted (e.g. the mechanics of Lorentz transformations which
is commonly known as special relativity; see for example \cite{SochiSpecRel,SochiBook4,SochiDingle2024}).
In fact, even the ``humble'' classical mechanics of the seventeenth
century is not well-understood at the epistemological level from various
aspects; the simple example is the interpretative meaning and significance
of Newton's third law despite its obvious simplicity and ``triviality''
from a formal and practical perspective.

Apart from these interpretative and epistemological difficulties and
the challenges they represent from this perspective, this inhomogeneity
of contemporary physics and the epistemological challenges it faces
(whether these epistemological challenges are the result of this inhomogeneity
or not) should put a question mark (at least from an epistemological
perspective) on the search for a unified physical theory (i.e. ``theory
of everything'') which is the ultimate dream of contemporary physicists.
In other words, is it possible or sensible or legitimate to have a
consistent unified physical theory when this unified theory is no
more than a ``stitching together'' of a collection of incompatible
theories related to separate (and sometime unrelated) phenomena? Moreover,
let's assume that we obtained such a theory by ``stitching together''
the branches of contemporary physics and hence we got a ``theory
of everything'' from a formal perspective, is this theory a ``theory
of everything'' from an epistemological perspective, i.e. can this
formal ``theory of everything'' provide an ``understanding of everything''
by providing a sensible and reliable ``interpretation of everything''.

The physical world is single and homogeneous and hence any ``theory
of everything'' that really reflects this single and homogeneous
world should be (at least for the purpose of being epistemologically
sensible and understandable) single and homogeneous not a ``stitching
together'' mosaic of theoretical structures that belong to different
historical eras where they depict and formalize separate physical
phenomena and aspects of the natural world.

In the present series of papers we intend to address these issues
by investigating and assessing the epistemology of contemporary physics
where in each paper we investigate one aspect (or branch or discipline)
of contemporary physics. To summarize, the main objectives of these
papers are:
\begin{enumerate}
\item Trying to dive deep into the epistemological significance and the
interpretative meaning of the main branches and aspects of contemporary
physics to have a better understanding of the current state of physics
from an epistemological and interpretative perspective. In other words,
we want to bypass the usual (or conventional) treatment and employment
of physics as a collection of formal recipes for ``doing business''
to reach their philosophical and epistemological significance and
spirit in a human-understandable language away from their formal symbolic
and perplexing forms.
\item Trying to investigate and assess the consistency of contemporary physics
from an epistemological perspective and hence investigate and assess
the merit of searching for a unified physical theory (i.e. ``theory
of everything'') from an interpretative and epistemological perspective
(if not from a formal and practical perspective).
\item Trying to suggest and propose alternatives to the current approaches
for tackling the subject of unification of physical theories which
have failed so far to produce a single unified physical theory (epistemologically
as well as formally) despite the huge efforts and despite the many
partial successes at formal level.
\end{enumerate}
\vspace{0.2cm}As indicated earlier, the current paper is the first
of a series of papers that we intend to release about the epistemology
of contemporary physics in its fundamental forms and formulations.
In this introductory paper we intend to lay down the foundations that
we need and will use in this investigation. In fact, the content of
this paper is partly based on our previous investigations about these
issues (see for instance \cite{SochiBook8,SochiKetab1}) and hence
the interested readers should refer to our previous books and papers
for more detailed treatments of these issues.

\section{\label{subObjectivesOfScience}The Objectives of Science}

There are two main objectives for science in general:

\noindent $\bullet$ \textbf{Practical  objective} which is conquering
the world and benefiting from its resources (or making use of it).

\noindent $\bullet$ \textbf{Theoretical objective} which is understanding
the world (or making sense of it).

\vspace{0.2cm}It is worth noting the following points:
\begin{enumerate}
\item These are the \textit{main} objectives for science\textit{ in general}
as opposite to individual branches of science which have more specific
and limited objectives (as well as these main and general objectives).
\item These two objectives are not limited to science but they generally
belong to all types of rational knowledge (whether scientific in a
technical sense or not).
\item These two objectives are not totally independent and hence better
understanding usually leads to better conquest while better conquest
generally leads to better understanding.
\item ``Understanding'' is an objective in itself (and hence it is as
important as ``conquest'') because we are intellectual species and
hence understanding is important to us like conquest although it may
not be regarded to be as important as conquest since it does not seem
to represent a direct biological need.
\end{enumerate}

\section{Formalism and Interpretation}

Formalism and interpretation are the two main components of any scientific
theory where the formalism largely reflects the practical aspect of
the theory (such as its applicability and predictivity) while the
interpretation reflects its human-understandable content and significance.
This means that the formalism of scientific theory is largely about
the practical objective of science while its interpretation is largely
about the theoretical objective (see $\S$ \ref{subObjectivesOfScience}).
Hence, a scientific theory is not complete \textit{in a strict sense}
unless it is consisting of these two components (e.g. we believe that
quantum mechanics is not a complete theory in a strict sense because
its formalism has no sensible interpretation in our view).

Broadly speaking, the formalism of a scientific theory is the theory
in its technical and rigorous scientific form (which is usually expressed
by well-defined terminology and symbols and cast into strict qualitative
or/and quantitative relationships which are usually in the form of
mathematical formulae such as equations and inequalities), while the
interpretation of a scientific theory is a rational explanation and
justification to the formalism that can be envisaged as a model that
we can understand and appreciate within our classic and macroscopic
experiences and expressed in our ordinary language and concepts.

In fact, the interpretation of a scientific theory may be seen as
the qualitative equivalent of the technical formalism of the theory
(which is usually quantitative in form). However, in our view the
interpretation is more than a qualitative reproduction of the technical
formalism because the role of interpretation is not just to provide
a more comprehensible and simple version of the theory (as represented
primarily by the formalism) but it extends beyond this to areas and
aspects like providing justifications, indicating causes and consequences,
and correlating the formalism to observable physical entities and
quantities.\footnote{In fact, one of the main roles and objectives of any interpretation
(according to the explicit meaning and suggestion of the word ``interpretation'')
should be providing a convincing explanation and justification to
the formalism.} Moreover, interpretation may not necessarily be a qualitative reproduction
of the formalism, as seen for example in the \textit{alleged} interpretations
of quantum mechanics where the formalism is based on obscure paradigms
like wavefunction which has no sensible qualitative equivalent (within
our classic macroscopic experiences whether practical or conceptual)
if we have to accept these as legitimate interpretations \textit{in
principle}.

Accordingly, the interpretation provides an added value to the original
theory (as represented primarily by the formalism) by vitalizing and
rationalizing it through describing and explaining its contents and
aspects using familiar concepts and understandable language and terminology
although this description and explanation may not necessarily be a
qualitative equivalent of its formalism if we have to avoid excluding
certain potential interpretations (although their eligibility may
be questionable).

It is worth noting the following points about the formalism and interpretation
of a scientific theory and the relationship between them:
\begin{enumerate}
\item The validity and invalidity of the formalism and interpretation are
generally independent of each other. Hence, we may have correct/incorrect
formalism with correct/incorrect interpretation. For example, we reject
special relativity as an incorrect interpretation (due to its axiomatic
and logical weaknesses) although the formalism of the theory (i.e.
the mechanics of Lorentz transformations) is generally supported by
strong experimental and observational evidence and hence it is correct
(i.e. in a practical, and possibly limited, sense). This similarly
applies to quantum mechanics in its experimentally-endorsed formalism
considering its various proposed interpretations which suffer, with
no exception, from major weaknesses and defects (as indicated earlier;
see $\S$ \ref{secIntroduction}).
\item We may have more than one acceptable or valid interpretation to a
given theory (as well as may have more than one theory or more than
one formalism that correctly describe a given physical phenomenon
(see for instance $\S$ \ref{subScienceNonUnique} and $\S$ \ref{secNatureKnow}).
\item We may not have an interpretation (or may not have an acceptable or
valid interpretation) of a particular scientific theory even though
the theory (as represented primarily by its formalism) is well established.
An obvious example of this case is quantum theory whose all existing
interpretations are questionable (in our view). In fact, some scientific
theories may not even be interpretable (such as quantum mechanics
in our view; see $\S$ 8.6 of \cite{SochiBook8}).
\item Ideal scientific theory is a theory whose formalism embeds its own
interpretation, i.e. the formalism intuitively suggests its interpretation
and hence the theory can be interpreted naturally and with minimal
effort.
\item In certain branches of physics, the difference between the formalism
and interpretation of scientific theory is rather delicate and subtle
(or blurred). This is especially true in those branches of modern
physics in which observation and measurement take a central role in
the theory itself (as it is the case with special relativity and quantum
mechanics). Anyway, it is very important to make the distinction between
formalism and interpretation clear and transparent to avoid traps
and pitfalls. For example, the confusion between formalism and interpretation
in Lorentz mechanics (where the interpretation of special relativity
is identified or mixed with the formalism of Lorentz mechanics) led
to many troubles such as giving the credit of the empirical success
of Lorentz mechanics to special relativity while putting the blame
for the epistemological failure of special relativity on Lorentz mechanics
(and hence we see the majority of physicists accept special relativity,
despite its epistemological inconsistencies, because of the empirical
success of Lorentz mechanics, while we see some physicists reject
Lorentz mechanics, despite its empirical success, because of the epistemological
failure of special relativity). Another example is from the quantum
theory where the status of the issue of measurement (and wavefunction
collapse which is one of the most troubling and problematic issues
in quantum mechanics) was confused, i.e. whether it belongs to the
formalism (as suggested by its inclusion in the axiomatic framework
of quantum mechanics according to its commonly-accepted version) or
it belongs to the interpretation (as suggested by its epistemological
nature and content).
\end{enumerate}

\section{\label{secLogic}Logic}

In simple terms, logic is a collection of rules and principles that
regulate our rational thinking to ensure consistency (or rather \textit{self-consistency})
and rationality. This means that the rules of logic should be respected
in any rational form of thinking and knowledge and this should obviously
include science both in its formalism and in its epistemology and
interpretation. So in brief, the compliance with logic in science
is an absolute necessity and hence any scientific theory that violates
logic (explicitly or implicitly, directly or indirectly, formally
or epistemologically, etc.) should be rejected without further ado
because this violation means that the theory is inconsistent and irrational
and hence it cannot be accepted as a form of rational knowledge.

Here we should point out (in reference to \textit{self-consistency})
that logic determines the rules of subjective consistency in the epistemic
system, while observation (in its comprehensive meaning that includes
experiment) determines the rules of objective consistency with the
external reality, meaning that the data and information obtained by
observation are consistent with (and within) the epistemic system.
In other words, logic ensures the internal consistency of the epistemic
system while observation ensures the external consistency of the epistemic
system, and thus logic and observation participate in ensuring the
validity and consistency of the epistemic system, both subjectively
(or internally or logically) and objectively (or externally or observationally).

So in brief, the basic principle upon which logic as a whole is based
is the principle of self-consistency which is no more than the principle
of non-contradiction in its comprehensive sense that includes similar
relationships (which may be called contrariety or likewise). Accordingly,
self-consistency is the essence and spirit of logic, and hence all
forms, patterns, rules, principles and so on in logic are no more
than details, examples and instances of the principle of self-consistency
and means for achieving it in specific forms, particular models and
different contexts. For example, the essence of syllogism (in its
various forms) is to ensure that this principle is observed in the
forms of inference and deduction in order to achieve consistency of
the result in itself, with its premises, and within the epistemic
structure on which the content of the syllogism is based.

The fact that the principle of self-consistency is nothing but the
principle of non-contradiction in its general sense and its most comprehensive
meaning plus the fact that self-consistency is the essence and spirit
of the entire logic are what give logic a special place and hence
protect it from skepticism and questioning and keep it away from disputes,
controversies and compromises that may extend to many other things
in any epistemic system (as indicated earlier in the first paragraph
of this section). For example, we may accept a weak form of realism
or causality (see $\S$ \ref{subPrinciplesOfReality} and $\S$ \ref{subPrincipleOfCausality2})
but we do not accept a weak form of logic.

We should finally note that the principle of self-consistency (upon
which logic in general is based and which is the basic building block
for establishing any epistemic system) is an instinctive principle
that is theoretically manifested among species which are high in the
rank of cognitive development. This principle, in essence, is instinctively
inherent in all living beings and manifests itself in different forms
in their behavior, actions and activities. In fact, it is the guarantee
of their existence and survival by meeting their needs, satisfying
their desires and enabling them to adapt to their environment. Without
this principle (i.e. in its instinctive form) the biological rules
and mechanisms that guarantee the survival and safety of the living
organisms will be violated and disrupted.\footnote{For example, without it, the prey may seek its predator, the danger
is prevented by what is worse, the thirst is quenched by what causes
thirst, the hunger is repulsed by abstaining from food or eating poison,
and so on.} This means that the principle of self-consistency (which is the essence
of logic) has biological instinctive roots and did not emerge from
nothing or originate from abstract conventions. In fact, this should
add more justification to our previous claims that the compliance
with logic is an absolute necessity and logic cannot be compromised
or have a weak form.

\section{The Epistemological Principles of Science}

In the following subsections we outline the principles that underlie
the epistemology of science and hence they provide the required foundations
for the development of scientific theories (and interpretations in
particular). However, it should be noted that these principles represent
our viewpoint; moreover we present only those principles that are
of primary interest and use to us in this investigation. We should
also note that these are epistemological principles for knowledge
in general and not only for science, i.e. science is just an instance
which we specify here because it is the subject of our investigation.

\subsection{\label{subPrinciplesOfReality}The Principles of Reality and Truth}

The epistemological foundations with regard to reality and truth are
summarized in the following three basic principles which are pivotal
not only to science but to all types of rational knowledge:

\noindent $\bullet$ The \textbf{existence of reality} which means
the existence of a real world beyond and outside the observer where
the reality of this world is independent of the observer.

\noindent $\bullet$ The \textbf{uniqueness of reality} which means
that this reality (as identified in the previous point) is unique
and hence we have only one reality.

\noindent $\bullet$ The \textbf{uniqueness of truth} which means
that we have only one truth which represents the honest reflection
of the (existing and unique) reality (as identified in the previous
points).

In the following points we discuss briefly some important issues related
to these principles:
\begin{enumerate}
\item Scientists should consider these principles from a purely epistemological
perspective rather than from an ontological perspective. Hence, the
existence of reality, for instance, should not mean the existence
of this ``alleged'' reality ontologically and in itself as an outside
entity but the existence of this alleged reality epistemologically
and for ourselves as an entity that we need to assume for building
and justifying our scientific knowledge. In fact, the ontological
reality (i.e. the ontological existence of reality) is a purely philosophical
issue and is irrelevant to science and out of its scope and domain.
\item The principles of reality and truth should be regarded as an epistemological
necessity because no rational and consistent science (and knowledge
in general) can be built without these principles (noting that they
are an ontological choice, i.e. we are free to accept or reject them
from an ontological perspective).
\item The principles of reality and truth do not mean that the world should
look the same to every species and individual. In particular, the
uniqueness of truth does not mean that the truth is absolutely definite
and determined. So, we can say the truth in its exact details is not
unique (although it is unique in its essence and within the given
conditions and circumstances). We may also say: the truth in its exact
details is unique only for a given individual considering the entire
set of conditions and considerations that determine the truth such
as time, location, measuring equipment and so on. Also see $\S$ \ref{subScienceNonUnique}
and $\S$ \ref{secNatureKnow}.
\end{enumerate}

\subsection{\label{subPrincipleOfCausality2}The Principle of Causality}

The essence of the principle of causality is the claim of an \textit{intrinsic}
association between a given event (or events) called cause and another
given event (or events) called effect (where we use ``claim'' to
indicate that causality cannot in general be proved, i.e. it is essentially
a postulate or a hypothesis which is ultimately based on intuition;
see $\S$ \ref{secIntuition}). This association supposedly makes
the occurrence of the effect inevitable when the cause occurs. In
fact, there are many aspects in the principle of causality that deserve
inspection and attention. However, due to the limited space and scope
of this paper we briefly discuss in the following points only those
aspects which are most relevant to our subsequent discussions and
investigations:\footnote{We note that further discussions and details about the principle of
causality can be found in \cite{SochiBook8,SochiKetab1}. Moreover,
other aspects of the principle of causality (which are more specific
to certain branches and aspects of physics) will be discussed in the
upcoming papers of this series.}
\begin{enumerate}
\item Similar to what we said in $\S$ \ref{subPrinciplesOfReality} about
the principles of reality and truth, scientists should consider the
principle of causality from a purely epistemological perspective rather
than from an ontological perspective.
\item In our view, the principle of causality is an epistemological necessity
for any rational theory (whether scientific or not) because without
this principle no rational and persistent relations and explanations
can be established (noting that this principle is a choice rather
than a necessity from an ontological perspective). So, the epistemological
demand for this principle is based on the epistemological demand for
rationality in science (and indeed in any type of rational knowledge)
as well as persistency which is a requirement for predictability.
Potential limitations (as well as other aspects within the limits
of size and scope of our investigation) of this principle will be
inspected and assessed in the following.
\item As suggested already, the justification for embracing the principle
of causality is its ability to provide a basis for predictability,
rationality, consistency, persistency and so on. These factors should
explain the need for the principle of causality as an epistemological
necessity regardless of accepting or rejecting the causality principle
at the ontological or philosophical level and regardless of its potential
limitations. So in brief, no science or rational knowledge can be
established without this principle in some shape and form although
the details of the extension and limitation of this principle and
its exact nature may be subject to debates and disputes.
\item A weak form of causality may be accepted (if we are forced to adopt
such a weak form) as long as rationality can be maintained.\footnote{We note that maintaining rationality with such a weak form may require
some modifications and adaptations to our basic conceptual framework
which basically rests on our classical intuition.} In fact, this should also apply to realism as represented basically
by the principles of reality and truth (see $\S$ \ref{subPrinciplesOfReality}),
i.e. a weak or partial form of realism may be accepted as long as
rationality can be maintained.\footnote{It is important to remember (see $\S$ \ref{secLogic}) that this
does not apply to logic (i.e. there is no acceptable weak form of
logic) because any violation of logic is a violation of consistency
and rationality which destroys the entire science and rational knowledge.}
\item One of the main aspects of the principle of causality is the chronology
of events in the causal relationships, i.e. the aforementioned \textit{intrinsic}
association is hierarchical in nature where the cause is supposed
to produce the effect in a certain chronological order between the
two, i.e. the cause and effect occur simultaneously or the effect
follows the cause in time. Accordingly, this hierarchy does not allow
the occurrence of the effect before the occurrence of the cause, i.e.
the effect cannot \textit{temporally} precede its cause. However,
we may find in modern physics (or rather in the literature of some
theories of modern physics) the so-called ``delayed cause'' relationships
where the effect supposedly precedes its cause (although the ``delayed
cause'' may not be really a cause and the relations may not be really
causal). Anyway, delayed cause should violate the causality principle
(and hence it should be rejected) if we accept the necessity of the
aforementioned chronological order in the causal relationships.
\item Any talk about causal relationships across different worlds (e.g.
macroscopic and microscopic worlds) whether in interaction or observation\footnote{``Interaction or observation'' refer to the cases where the cause
and effect are interacting in the same world or interacting across
different worlds (i.e. one of them is in a world and the other is
in another world) while the observer is in a world different from
the world of both or one of them. This applies to all our physical
observations (as well as theories, knowledge and so on) regarding
the quantum and astronomical and cosmological worlds since we are
classical (or macroscopic) in scale and hence we (as observers) cannot
be ``quantum observers'' or ``astronomical observers'' or ``cosmological
observers'' (e.g. when we observe a quantum phenomenon we observe
it from our classical macroscopic world since we cannot penetrate
the quantum world by becoming quantum objects or quantum creatures).
In fact, we should accept the fact that we are trapped in one world
and separated by an impenetrable barrier that detaches us from all
other worlds (so all our observations to phenomena belonging to other
worlds are actually indirect in this sense). See also $\S$ \ref{secIntuition}
and $\S$ \ref{subScale}.} should address the following three issues (among other issues):\\
$\bullet$ The existence or not of absolute global time.\\
$\bullet$ The meaningfulness of a unified concept or definition of
``time'' across all worlds.\\
$\bullet$ The unified reality of ``time'' across all worlds.\\
These issues should be naturally manifested, for instance, when dealing
with the alleged ``delayed cause'' where these causal relationships
involve occurrence or/and observation across different worlds.
\item Regarding the justification of the aforementioned \textit{intrinsic}
nature of the causal association, the internal sense of ``constructional''
relationship between the cause and the effect is what justifies this
intrinsic nature. In other words, the repetitive association in itself
(even if it is perpetual) does not imply causality unless we have
a ``sense'' or ``feeling'' or ``intuition'' of a constructional
and inherent relationship between the cause and the effect (where
this feeling normally originates from our intuition which is generally
based on our past experiences). In fact, this is inline with what
we indicated earlier that causality is essentially postulated rather
than proved. Accordingly, association in itself (even when it is repetitive
and perpetual) in not sufficient to establish causal relationship.
On the other hand, we may be justified to conclude a causal relationship
between two events from a single observation based on our sense or
feeling (or rather intuition) of this inherent relationship. This
should indicate that even if we believe that the origin of the principle
of causality is induction (and empiricism in general which is based
on direct observation), the specific causal relationships require
more than induction (or observation) since they are based on our internal
sense and rational thinking that is based on our overall past experiences
and knowledge (i.e. these relations require deduction and intuition
as much as they require induction and observation). So in brief, perpetual
association may not imply causality while casual association (or even
hypothesized association) may imply causality where the ultimate criterion
and origin of this implication stem from our intuition (see $\S$
\ref{secIntuition}) which is largely based on our past experiences
and the presumed epistemic system (which is generally and ultimately
based on our past experiences as well as our cognitive capabilities
as intellectual species).
\item As indicated earlier, the causality relationship at the empirical
level is no more than an association (continuously-repeated in the
past and supposedly continuously-repeated in the future) between two
events (or sets of events). The distinction between the cause and
the effect in this relationship is decided either by the chronology
of the events (i.e. the first-occurring is the cause and the second-occurring
is the effect) or by the dependency of the events (i.e. the independently-occurring
is the cause and the dependently-occurring is the effect). However,
we should note that chronology and dependency may not be available
in some causal relationships (e.g. simultaneous events not under our
control) and hence we should rely on other means (mainly intuition)
to distinguish between cause and effect. This should endorse what
was indicated earlier that empiricism alone is not sufficient to establish
causality relationships in general.
\item Whether the principle of causality is restricted to the classical
(or macroscopic) world or it is valid even in the non-classical worlds
(like quantum and astronomical and cosmological worlds) seems to be
a matter that can be debated and disputed. However, as a requirement
of rationality and consistency the principle of causality should be
general and hence it should apply to the non-classical worlds as to
the classical world. Yes, the specific causal relationships may be
less obvious with regard to the non-classical worlds since we have
no familiar experience or direct observation or justifiable intuition
about the non-classical worlds and their phenomena and their relationships.
Nevertheless, we should note that most or all of the specific causal
relations of the non-classical worlds are observed and deduced indirectly
through classical macroscopic phenomena (represented typically by
the reactions of the measuring devices which are generally classical
macroscopic let alone our senses and cognitive capabilities which
are obviously classical macroscopic) and hence the classical macroscopic
relationships should be rationally projected onto the non-classical
relationships. Yes, an issue may be raised about the extent of the
validity of the principle of causality and possible restrictions on
it at the non-classical level (to justify for instance the seemingly
non-causal features of quantum behavior or the non-causal nature of
the origin of Universe whether we believe in a deity or not).
\end{enumerate}

\subsection{\label{subScienceNonUnique}The Principle of Non-Uniqueness of Science}

In our view, science is not unique and hence in principle any physical
phenomenon can be described, quantified and predicted correctly (i.e.
without violation to the rules of logic or the principles of reality
and truth; see $\S$ \ref{secLogic} and $\S$ \ref{subPrinciplesOfReality})
by more than one scientific theory.\footnote{In fact, this principle in its general form is a principle of non-uniqueness
of knowledge (i.e. rational knowledge). So, science (which is the
primary interest in this investigation) is just an instance of knowledge.} Hence, any scientific theory should be replaceable (at least in principle)
by another scientific theory where both theories are empirically correct
and practically equivalent although they may be epistemologically
different (in addition to their difference in formalism as well as
potential difference in merit and advantages/disadvantages).

This principle is very important to science since it allows and legitimizes
this sort of diversity in science which is not obvious and seems to
be unrecognized among scientists and scholars. In fact, we feel that
there is a common undeclared belief or consensus that science is unique
and hence new theories are legitimate to emerge only as corrections
or improvements (e.g. by generalization) or as minor and superficial
modifications of old theories (as long as old theories are working
in general). In fact, there are many examples and instances in science
(e.g. in classical and quantum mechanics) that justify and legitimize
our adoption of the principle of non-uniqueness of science.

\subsection{\label{subEconomy}The Principle of Economy}

The essence of this principle, which may also be labeled with other
tags like ``Occam's razor'' or the ``law of parsimony'', is that
the scientific theory should be as simple as possible, and hence if
we have a set of theories (or formulations or interpretations, ...
etc.) that are equivalent in their predictions and outcomes then we
should choose the simplest one.

It is worth noting the following points about this principle:
\begin{enumerate}
\item The principle of economy (in some of its instances and interpretations)
should imply the validity of the principle of non-uniqueness of science
(see $\S$ \ref{subScienceNonUnique}).
\item The principle of economy does not necessarily require selection between
different theories but it can be used in the creation or emergence
of a single theory where economy considerations (e.g. in assumptions
and postulates) are taken into account in its creation and formulation.
\item This principle does not represent a necessity or obligation and hence
it can be violated (although this should be for good reason).
\item This principle should not only be used for the selection or creation
of theories but it should also be used for the application of theories.
In fact, the latter use is a common practice (and hence it gives more
legitimacy and extent to this principle).
\end{enumerate}

\subsection{\label{subPrincipleOfIntuitivity}The Principle of Intuitivity}

The essence of this principle is that certain theories (or theses
or opinions or $\ldots$) are intuitive because they comply with our
internal sense which is largely based on our past experiences, while
others are not, and hence we can accept and reject certain theories
or make preferences among them according to this intuitivity criterion
(i.e. intuitive or more intuitive theories are accepted or preferred
and vice versa). In fact, terms like ``intuitive'' and ``counter-intuitive''
are common in the literature of science when it comes to assessing
and accepting or rejecting certain theories and opinions. This means
that the use of intuitivity criterion for assessing and selecting
theories in science is legitimate as a basis for making scientific
judgments and preferences and is generally acceptable among the scientific
community.

It is worth noting the following points about the principle of intuitivity:
\begin{enumerate}
\item The justification of this principle is essentially based on the theoretical
objective of science as an endeavor for understanding the world and
making sense of it (see $\S$ \ref{subObjectivesOfScience}) noting
that intuition is one of the main sources and manifestations of understanding.
\item The principle of intuitivity is generally not compulsory and hence
it is mostly used in making preferences and voluntary choices (and
so it is like the principle of economy in this regard; see $\S$ \ref{subEconomy}).
\item The principle of intuitivity is mostly related to the interpretation
of theories (due to the link of this principle to the ``understanding
objective'' of science which is the essence of interpretation; see
$\S$ \ref{subObjectivesOfScience}) although it may also be used
for other purposes and in other contexts.
\item We must distinguish between intuition as an active element in the
cognitive processes (see $\S$ \ref{secIntuition}) and the principle
of intuitivity as an epistemological principle (which is the subject
of the current subsection). While the first is rooted in the cognitive
processes and is indispensable to them, the second is generally optional
and not obligatory, and thus it is often used to select preferences
and choose favorable options (as discussed earlier).
\end{enumerate}
\vspace{0.2cm}Also see $\S$ \ref{secIntuition}.

\section{\label{secIntuition}Intuition}

In $\S$ \ref{subPrincipleOfIntuitivity} we discussed the principle
of intuitivity and distinguished between intuition (or rather intuitivity)
as an epistemological principle and intuition as an active element
in the cognitive processes. In fact, intuition as an active element
is the essence of our topic in this section, in which we address the
role of intuition in the cognitive processes in general and its crucial
contribution to the creation, development and maintenance of epistemic
systems and structures.

In summary, intuition provides a mechanism for creating and synthesizing
novel concepts (ideas, relationships, etc.) based on previous experiences
and adapting available stereotypical concepts that are flexible and
capable of modification as well as using and applying previously obtained
concepts, in the face of subjective epistemic experiences and objective
observations with the aim of integrating them consistently and rationally
into the existing epistemic system or employing them beneficially
within it.

By doing so, intuition actually ensures not only the enrichment of
the existing epistemic system (by addition, generalization, extension,
and so on), but also solving theoretical and practical problems for
the system which cannot be solved without the intuition. In fact,
intuition in the intelligent species plays a distinct and special
role in achieving optimal adaptation and best satisfaction through
enriching the epistemic system as well as resolving its complications
and dilemmas and correcting its defects and faults. In brief, intuition
plays a prominent role in building the epistemic system, rationalizing
it, justifying it, complementing it, correcting its defects and shortcomings,
filling its gaps, and so on.\footnote{Intuition is actually the essence of intelligence. In fact, the essence
of artificial intelligence in our view is the formation of an ``introspective
intuition'' for the intelligent computer by teaching it to benefit
from past experiences through training and learning to extract models
and patterns that are suitable for use in similar circumstances and
situations by benefiting from collecting, classifying and analyzing
the available information and data.}

We should finally note that our intuition is created and formed by
our classical macroscopic experiences and hence we do not have reliable
intuition (since we have no direct experiences and direct observations)
with regard to other worlds (e.g. quantum and cosmological worlds).
Accordingly, we cannot make reliable judgments based on our intuition
about other worlds since our intuition is ``classical macroscopic''
while the other worlds are not (see $\S$ \ref{subScale}).

\section{\label{secNatureKnow}The Nature of Human Knowledge}

We discuss in the following subsections a number of general aspects
about the nature of human knowledge (represented mainly by science
inline with our scope and perspective) which are important to be aware
of.

\subsection{\label{subHumanism}Humanism}

Science is a product of mankind and hence it is characterized by the
features of our cognitive system and how we think and interact with
our environment (i.e. Nature). In fact, ``humanism'' does not only
represent the ``species'' factor that enters in the determination
and identification of the scientific process and knowledge, but it
also includes many other factors such as cultural, social and personal
factors. This, for instance, could partly explain the fact that we
can have more than one correct scientific theory for describing and
formulating a single physical phenomenon. As we saw earlier (refer
for instance to $\S$ \ref{subScienceNonUnique}), any physical phenomenon
can be correctly described and formulated by more than one scientific
theory without violating the rules of logic or the principles of reality
and truth (see $\S$ \ref{secLogic} and $\S$ \ref{subPrinciplesOfReality}).
This is because any type of human knowledge is not really an ideal
image-reflection (or image extraction) of reality but it is rather
an interaction between the thinker and his environment (i.e. Nature)
and hence this process (as well as its outcome which is the ``extracted
image'') is affected by many humanist factors. In other words, the
acquired knowledge (or the ``extracted image'') is actually an artistic
impression of the thinker by Nature, i.e. it is more like an artistic
impressionist painting than a high-definition photograph or mirror-reflection.

So in brief, correct science does not represent a pure image or mirror-reflection
of reality but it is a mix of reflection (or discovery) and invention
(or creation) where this process and its outcome are affected and
determined by many humanist factors (in the broadest sense of ``humanist''
as indicated above).

\subsection{\label{subScale}Scale of Observer}

This factor may be seen as originating from the previous factor (see
$\S$ \ref{subHumanism}) although we consider it separately due to
its specificity and particular importance to science. The essence
of this factor is that because we are a species of a given size (or
rather scale), our sensory and conceptual experiences are acquired
from the natural world (as perceived by us) on the scale that is comparable
to our size (and in fact it is processed by our ``humanist'' nature
which is the product of an evolutionary process on this scale and
size). This means that the validity of the patterns (models, concepts,
paradigms, etc.) that we acquire from our past physical experiences
and we use as elements and prototypes in our scientific theories is
generally restricted to certain scale and size and hence these patterns
and models are not necessarily valid as physical models to the world
on other scales and sizes.

This is based on the fact that the patterns and models (that are acquired
and developed as abstractions from the physical world) are generally
scale-dependent and hence some patterns and models may be compatible
with certain scales but not with other scales. In other words, not
all patterns and models are valid for any scale and size (or not all
patterns and models are valid for describing the physical world at
any scale and size). For example, the physical pattern or model of
``wave'' is acquired from our physical experiences at classical
(or macroscopic) scale and hence the ``wave'' model may not be sensible
to describe physical objects and phenomena at pico or femto scales
because these scales were beyond  our reach during the acquisition
and development of the ``wave'' model and hence this model may not
reflect the nature of the world at these scales. This similarly applies
to many other concepts, models and patterns like ``color'' or ``brittleness''.\footnote{In fact, if the invalidity of ``wave'' at pico or femto scales is
suspicious the invalidity of ``color'' or ``brittleness'' (and
their alike) should be obvious because they are fundamentally macroscopic.
For instance, having red and green electrons or brittle and ductile
atoms is obviously nonsensical.}

In the following points we provide more clarifications about the ``scale
of observer'' as an important factor that determines the nature of
human knowledge (and hence as a factor that should be considered in
the assessment and validation of scientific models and theories):
\begin{enumerate}
\item ``Scale'' is more general than ``size'' and hence it includes
for instance speed (i.e. how fast or slow) and intensity (i.e. how
strong or weak a certain physical property like electric charge or
gravitational field). So in brief, ``scale'' is about the magnitude
of a physical attribute (whether size or something else).
\item ``Scale'' is what determines the nature and essence of our intuition
(see $\S$ \ref{secIntuition}) . Hence, considering ``scale'' factor
our intuition is classical macroscopic because we are classical macroscopic
creatures and hence our scale as observers which determines the nature
and essence of our intuition is also classical macroscopic. This means
that we can legitimately use our intuition only with regard to the
classical macroscopic world not other worlds (i.e. quantum or astronomical
or cosmological worlds).
\item The ``scale'' factor is actually about the \textit{relative} magnitude
as determined by the scale of the observer in comparison to the scale
of the observed phenomenon. This is because the scale of the observer
should primarily determine the scale of the observed phenomenon to
which the patterns and models of the observer (as well as his intuition)
are appropriate or not.
\item \label{enuQuantumSize}The ``scale'' factor is what makes quantum
physics and Lorentz mechanics special among scientific branches and
theories. This is because the extreme smallness of quantum objects
and the extreme fastness of Lorentzian objects make them unusual and
hence our physical models and patterns as well as our intuition (which
were developed and acquired during our long evolutionary history dealing
with macroscopic and slow objects) are not suitable to describe and
quantify quantum and Lorentzian phenomena in a sensible and interpretable
way (or at least this requires exceptional effort and attention and
hence ignoring this fact leads to the complications and contradictions
which are associated with quantum physics and Lorentz mechanics).
In fact, we believe that this should similarly apply to other scientific
theories and branches such as astronomy, astrophysics and cosmology
where the scale in space and time (as well as other properties and
attributes such as intensity of electromagnetic or gravitational fields)
is unusual and hence it should not be considered or treated as classical
or macroscopic. Accordingly, we may need to revise our principles,
laws, assumptions, concepts, models, etc. that we are currently using
in the investigation of subjects like astronomy and cosmology because
these classical macroscopic principles (... etc.) may not be appropriate
to use due to the scale factor. In fact, the failures in these fields
(even by some supposedly ``non-classical'' theories like general
relativity) could be a sign for the failure of our classical macroscopic
principles (... etc.) in these fields because of the scale factor.\footnote{These failures are demonstrated, for instance, by the need for dark
energy or dark matter or creation or nonsensical consequences like
travel in time ... etc. which modern physics (in these subjects in
particular) is full of.} These issues will be pursued in the forthcoming papers of this series.
\item The scale factor could have an effect even on some of our most fundamental
and seemingly-intuitive concepts like space and time or our philosophical
and epistemological patterns and paradigms such as our concept about
physical reality and its nature. For example, the paradigms of ``space''
and ``time'' may not be applicable or appropriate (at least in their
exact sense) at very large or very small scales (such as the scales
of Universe and nucleon). Similarly, ``macroscopic realism'' may
not be exactly applicable to ``microscopic realism'' (because ``macroscopic
reality'' may not be identical to ``microscopic reality''). In
fact, even the epistemology can be scale dependent (especially in
its detailed interpretative aspects).
\item Can the scale factor affect logic (and hence we may have modified
or different versions of logic for worlds and realities of different
scales such as quantum logic for the quantum world and cosmological
logic for the cosmological world)? In fact, the invention (or discovery)
of a new type of logic (which supposedly departs from the ordinary
logic) that is valid to worlds at scales different to the ``classical
macroscopic scale of the ordinary logic'' may be seen to be as legitimate
as the invention of the non-Euclidean geometries (which depart from
the Euclidean geometry). However, the ``conceptual legitimacy''
of any type of non-Euclidean geometry is based on its self-consistency
(which is the essence of the ordinary logic as indicated earlier;
see $\S$ \ref{secLogic}) while its ``practical legitimacy'' is
based on its agreement with physical observations (as well as its
usefulness in this regard), i.e. external consistency (see $\S$ \ref{secLogic}).
Similarly, the ``conceptual legitimacy'' of any type of novel logic
(such as quantum logic) should be based on its self-consistency (which
originates from the ordinary logic) while its ``practical legitimacy''
is based on its agreement with ``observations'' (and its usefulness
in this regard). This means that the legitimacy of any novel logic
should be (at least partly) acquired from the ordinary logic (and
hence to a certain extent it should be consistent with the ordinary
logic). After all, we are classical macroscopic creatures (rather
than quantum or cosmological creatures for instance) and hence even
``quantum or cosmological logic'' (i.e. the logic that belongs to
the quantum world or cosmological world assuming the existence of
such logic) should be subject to the rules of our ``classical logic''
(i.e. ordinary logic) to be ``sensible'' and ``logical'' to us.
So in brief, at this stage we assume there is no scale-dependent logic
(e.g. ``quantum logic'' or ``cosmological logic'') as an alternative
and substitute to the ordinary logic. Yes, there may exist some logical
rules that are tailored specifically to worlds other than our classical
macroscopic world (e.g. quantum and cosmological worlds) and they
are consistent with the rules of ordinary logic. In this case, they
are just instances of the rules of ordinary logic and hence they should
be acceptable (although they are not expected to be of general validity
like the rules of ordinary logic). So in brief, logic by nature is
``classical'' but scale-independent because we are classical creatures
and logic is about our own consistency (or ``self-consistency'')
and hence it cannot be changed in a fundamental way because we cannot
be other than ourselves.
\item The aforementioned fact that our patterns and models, as well as our
intuition, are scale-dependent should lead to an obvious intrinsic
limitation in our ability to describe and interpret (in a realistic
way) physical phenomena incommensurate to our scale. In fact, there
are many examples in physics about this limitation and how we use
(justifiably or unjustifiably) scale-limited patterns and models approximately
or inappropriately to describe and interpret phenomena that are beyond
our scale because we have no other (more realistic) choice. In fact,
this highlights an essential limitation of science in general. After
all, we are \textit{humans} (see $\S$ \ref{subHumanism}) and of
certain physical \textit{scale} and hence we have no direct access
to entities that are beyond our familiar experiences which are acquired
and shaped according to our type and scale (and which generally form
and shape our intuition; see $\S$ \ref{secIntuition}). Therefore,
we may be content to use these scale-limited models and patterns (but
cautiously) as approximate prototypes to investigate these entities.
The danger, however, emerges when these models and patterns are treated
(recklessly) as realistic and exact prototypes (and this sort of recklessness
seems to be common in science as in all other aspects of human life).
\end{enumerate}

\subsection{The Evolution of Human Knowledge}

We present in the following points a number of general notes about
the evolution of human knowledge (represented mainly by science inline
with our scope and perspective) which are important to be aware of:
\begin{enumerate}
\item The evolution of knowledge is probabilistic in nature (i.e. it is
not deterministic). This means that the development or evolution of
knowledge has an infinite number of probabilistic paths and ways,
and in taking this or that path, it passes through crossroads whose
choices and outcomes are determined by probabilistic and non-deterministic
factors such as the historical and geographical conditions and circumstances,
the prevailing cultures, the neighboring and intersecting epistemic
trends and currents, the personal characteristics and individual traits
of the elite thinkers and influencers (such as scientists, philosophers,
and even political leaders) as well as countless other factors. So,
knowledge can probabilistically follow one path or another where different
paths vary in their integrity, virtues, vices, inputs, and outputs.
For example, a theory or an idea or a political event or a distinguished
individual (with his personal characteristics) can contribute to the
path of knowledge and its development positively or negatively (or
positively in some aspects and negatively in other aspects, which
is often the case) and may take it along one route or another. In
brief, knowledge is a historical process and a probabilistic event
that is not predetermined by necessities other than the necessities
of the status quo, which is subject to countless overlapping and intertwined
probabilistic factors. In fact, the process of knowledge creation
and knowledge evolution is subject to the same rules that govern the
emergence and evolution of living species, and hence these rules should
explain the creation and evolution of different knowledge and epistemic
systems just as they explain the emergence and evolution of different
living species. It is important to note that this is one (and perhaps
the most important) of the origins, causes, reasons and justifications
for the principle of non-uniqueness of knowledge (and science in particular;
see $\S$ \ref{subScienceNonUnique}).
\item Although the general trend in the evolution of knowledge is progress
and advance, it may fluctuate and even go backwards from time to time.
So, it is not necessarily that the present knowledge is better than
the past knowledge or the future knowledge is better than the present
knowledge. In brief, knowledge is integrative and progressive in
the long term, taking into account short-term fluctuations, and hence
although the general trend in the evolution of knowledge is to progress
and move forward, it may stagnate, fluctuate, and even go backwards
from time to time.
\item A theory or an idea or a development in the history of science may
contribute positively in some aspects and negatively in other aspects.
For example, modern physics (despite its undeniable empirical successes
and achievements) is not necessarily a progress and success in its
entirety (e.g. it can be a setback from certain epistemological perspectives).
\end{enumerate}
\pagebreak{}

\phantomsection 
\addcontentsline{toc}{section}{References}\bibliographystyle{plain}
\bibliography{Bibl}

\end{document}